\documentclass[12pt, preprint]{aastex}
\usepackage{amsmath}
\usepackage{graphicx}
\newcommand{\fdegr}{.\!\!^\circ}

\newcommand{\msun}{\rm M_{\sun}}
\newcommand{\Msun}{\rm M_{\sun}}

\newcommand{\rchinu}{\chi^{2}/\nu}

\newcommand{\pchinu}{\chi_\wp^{2}/\nu}

\newcommand{\kpc}{\rm kpc}

\newcommand{\cm}{\rm cm}

\newcommand{\spin}{a_{*}}

\newcommand{\chandra}{{\it Chandra~}}

\shorttitle{Jet Kinematics in J1550}
\shortauthors{Steiner \& McClintock}

\begin{document}

\title{Modeling the Jet Kinematics of the Black Hole Microquasar\\ XTE J1550-564:  A Constraint on Spin-Orbit Alignment}

\author{James F.\ Steiner\altaffilmark{1} and Jeffrey E.\ McClintock\altaffilmark{1}}

\altaffiltext{1}{Harvard-Smithsonian Center for Astrophysics, 60
  Garden Street, Cambridge, MA 02138.} 
\email{jsteiner@cfa.harvard.edu}

\begin{abstract}

  Measurements of black hole spin made using the continuum-fitting
  method rely on the assumption that the inclination of the black
  hole's spin axis to our line of sight is the same as the orbital
  inclination angle $i$ of the host binary system.  The X-ray and
  radio jet data available for the microquasar XTE J1550--564 offer a
  rare opportunity to test this assumption.  Following the work of
  others, we have modeled these data and thereby determined the
  inclination angle $\theta$ of the jet axis, which is presumed to be
  aligned with the black hole's spin axis.  We find $\theta \approx 71
  \degr$ and place an upper limit on the difference between the spin
  and orbital inclinations of $|\theta - i| < 12$~deg (90\%
  confidence).  Our measurement tests for misalignment along the line
  of sight while providing no constraint perpendicular to this plane.
  Our constraint on the misalignment angle supports the prediction
  that the spinning black hole in XTE J1550--564 has aligned itself
  with the orbital plane and provides support for the measurement of
  its spin via the continuum-fitting method.  Our conclusions are
  based on a simple and reasonable model of a pair of symmetric jets
  propagating into a low density cavity whose western wall is
  $\approx20$\% closer to XTE J1550--564 than its eastern wall.


\end{abstract}

\keywords{black hole physics --- stars: individual (\object{XTE
    J1550--564}) --- X-rays: binaries}

\section{Introduction}\label{section:Intro}

Although it is thought that the Galaxy is host to tens of millions of
stellar-mass black holes, only about 50 have been discovered
\citep{Ozel_2010}.  All of them are accretion-powered X-ray sources
that are located in X-ray binary systems.  Most such systems, which
are similar to our featured black hole binary XTE J1550--564, have
short orbital periods ($P \sim 1~{\rm d}$) and are comprised of a
low-mass ($\lesssim 1~\msun$) donor star and a $\sim10~\msun$ black
hole.  A stream of gas from the Roche-lobe-filling star feeds into the
outer part of an accretion disk that encircles the black hole.  On a
time scale of weeks, viscous forces in the disk cause this gas to move
radially inward to the center.  Within a few hundred kilometers of the
black hole, the optically-thick gas reaches a temperature of
$\sim10^7$~K and produces an X-ray luminosity that is near the
Eddington limit ($L \sim 10^{39}$~erg~s$^{-1}$).  Accretion onto the
black hole is not a steady process: A typical source is luminous for
only about a year, and then it fades into a quiescent state for years
or decades.

XTE J1550--564 (hereafter J1550) is a much-studied Galactic black-hole
transient system that was discovered on 1998 September 6 using the
All-Sky Monitor (ASM) onboard the {\it Rossi X-ray Timing Explorer}
({\it RXTE}).  Thereafter, it was observed almost daily during its
entire 8-month outburst cycle using {RXTE's} pointed instruments
\citep{Sobczak_2000}.  Two weeks into outburst, the source abruptly
rose fourfold in intensity and produced a brilliant 7-Crab flare.
During this X-ray flare, J1550 was approximately at its Eddington
limit for $\approx1$~day \citep{Steiner_j1550spin_2011}.  Four days
later, radio observations made using the Australian Long Baseline
Array (LBA) revealed relativistic ejecta moving both eastward and
westward from J1550 \citep{Hannikainen_2009}.  The two components were
observed to be separated by $\sim250$ mas and moving at relative speed
of $\mu_{\rm app}\approx65$ mas/d, equivalent to an apparent
separation velocity of $\sim1.7$c.  Nearly two years later, \chandra
imaging observations revealed large-scale ($\gtrsim 20\arcsec$)
relativistic jets undergoing deceleration \citep{Corbel_2002}.  This
landmark discovery of a pair of ballistic X-ray jets was the first
detection of its kind for a Galactic source.

By modeling an extensive collection of optical and infrared data for
J1550, \citet{Orosz_Steiner_2011} have determined the mass of the
black hole primary, $M=9.1\pm0.6~\msun$, the distance to the
binary, $D=4.38^{+0.58}_{-0.41}~\kpc$, and the inclination of its
orbital plane, $i=74\fdegr7\pm3\fdegr8$.  Assuming that the black
hole's spin is aligned with the orbital angular momentum,
\citet{Steiner_j1550spin_2011} have measured the spin using the
continuum-fitting method to be $\spin = 0.34^{+0.20}_{-0.28}$, where
$\spin \equiv cJ_{\rm spin}/GM^2$ is the black hole's dimensionless
spin parameter and $J_{\rm spin}$ its angular momentum.  Steiner et
al.\ also measured the spin using the independent Fe-line method and
find $\spin = 0.55^{+0.10}_{-0.15}$; taken together, the two
measurements imply $\spin \approx 0.5$.  The continuum-fitting method
relies on a model for the thermal emission from an accretion disk
\citep{Zhang_1997}, while the Fe-line method relies on a model of the
relativistically broadened fluorescence features emitted by the disk
\citep{Fabian_1989}.

For a black-hole binary system like J1550, with a low-mass companion,
the ratio of the orbital angular momentum to the spin angular momentum
of the black hole is given by
\begin{equation}
J_{\rm orb}/J_{\rm spin} \approx 65~ \spin^{-1} \left(\frac{M}{10~\Msun}\right)^{-4/3}\left(\frac{M_2}{\Msun}\right)\left(\frac{P}{1~{\rm d}}\right)^{1/3},
\end{equation}
where $M_2$ is the mass of the secondary star.  For J1550, this ratio
is $\approx$50, and thus it is reasonable to expect that, given a
means of interaction, the spin of the black hole will eventually come
into alignment with the orbital angular momentum.  The time scale for
this to occur is an important question for continuum-fitting spin
measurements because in applying this method one generally must assume
that the two vectors are aligned.

If there is an initial misalignment between the spin and the orbital
angular momenta, then Lense-Thirring precession will cause the inner
X-ray-emitting portion of the disk to line up with the spin of the
black hole \citep{Bardeen_Petterson}.  At the same time, at very large
scales, the disk will align itself with the orbital plane, and the
transition between these regimes will manifest as a warp in the disk.
When a misalignment is present, the black hole will be torqued into
alignment by the accreting matter acting with a lever arm of order the
warp radius (e.g., \citealt{Natarajan_Pringle_1998}).  Using a
maximally conservative (minimum-torque) assumption,
\citet{Fragos_2010} concluded (based on a population synthesis study)
that the spin axes of most black hole primaries will be tilted less
than 10$\degr$.  Fragos et al. assumed that the torque acts at the
innermost stable circular orbit, $R_{\rm ISCO}<6 GM/c^2$ for $a_*>0$,
whereas the warp radius has been estimated to be located at $R_{\rm w}
\approx 200 GM/c^2$ \citep{King_2005, Lodato_Pringle_2006}.

For a typical system, the time scale for accretion to torque the black
hole into alignment has been estimated to be $t_{\rm align} \sim
10^{6}-10^{8}$ years
(\citealt{Martin_j1655_2008,Maccarone_2002})\footnote{\citet{Maccarone_2002}
overestimated $t_{\rm align}$ as the result of a numerical error in
his Eqn.~6, which implies a time scale that is 50 times longer than
that implied by his Eqn.~1.}.  Therefore, one expects alignment to
occur early in the lifetime of an old-population transient system,
such as J1550, and that most such systems will presently be well
aligned.

It is obviously important to test this theoretical expectation.
However, it has proved challenging to obtain a firm measurement of the
degree of alignment for any black hole binary.  Such a measurement
requires a determination of the position angle of the binary on the
plane of the sky.  While this may be possible in the future for J1550,
we lack the requisite orbital astrometric data for the system and are
therefore limited to testing for alignment along the line of sight.
Measuring the orbital inclination angle of the binary is relatively
simple and is routinely done by modeling optical data (e.g.,
\citealt{Orosz_2009}).  In contrast, it has proved difficult to obtain
reliable estimates of the inclination of the inner disk.

Currently, the most direct way of determining the inner-disk
inclination is by modeling jet ejecta, which are presumed to be
aligned with the black hole's spin axis.  For the case of symmetric
ejecta, see the review by \citet{Mirabel_Rodriguez}.  The jet ejecta
that are relevant to this paper are pairs of discrete, detectable
condensations of radio-emitting plasma, which we generally refer to in
shorthand as ``jets.''

An alternative approach to measuring the inner-disk inclination is via
the same Fe-line method that is used to measure black hole spin
\citep{Reynolds_Nowak_2003}.  However, existing models make
simplifying assumptions concerning how the ionization state of the
disk varies with radius.  Given that there is a degeneracy between
ionization and inclination in Fe-line/reflection models for
stellar-mass black holes, these inclination estimates are subject to a
systematic uncertainty of unknown magnitude.  Meanwhile, prospects are
good that more advanced reflection models will provide robust
estimates of inclination.

Based on observations of radio jets, two confirmed black hole systems,
GRO J1655--40 and SAX 1819--2525, are good candidates for hosting
misaligned black holes.  In the case of GRO J1655--40, using a
kinematic model for the jets and measurements of proper motion,
\citet{Hjellming_Rupen} reported a jet inclination angle of 85$\degr$.
However, the authors give no error estimate for either the jet
inclination angle or the proper motion.  Furthermore, the reliability
of the estimate for the jet inclination angle is called into question
by the intrinsic and variable asymmetries that were observed for the
opposing jets \citep{Mirabel_Rodriguez}.  Taking the 85$\degr$ jet
inclination angle at face value, one concludes that the jet axis and
orbital vector are misaligned by $>15\degr$ \citep{Greene_2001}.

In the case of SAX J1819--2525, the evidence is less certain.  There
is only a single observation of extended radio emission (because the
source faded promptly).  By making the assumption that this emission
was associated with a major X-ray outburst that occurred hours
earlier, superluminal motion ($\beta_{\rm app} > 10c$) and a
misalignment angle of $> 50\degr$ were inferred
\citep{Orosz_v4641_2001, Hjellming_2000}.  However, as
\citet{Chaty_2003} have argued, the jet may have been ejected a couple
of weeks before the major outburst, in which case the Lorentz factor
of the jet was modest and its inclination consistent with the
inclination of the binary.  This is a reasonable possibility given
that the source was observed to be active at optical wavelengths for
several weeks before the X-ray outburst.

Compared to the jets in these two systems and those in other Galactic
microquasars, the jet ejections observed for J1550 are remarkable.
They were observable for years (rather than weeks or months), and
therefore their physical separation from J1550 was observed to become
exceptionally large.  These are possibly the largest resolved jets
observed for any black hole when considering the dimensionless
distance between them, i.e., $d/M$ (\citealt{Hao_Zhang_2009}, and see
\citealt{Heinz_2002}).  By this measure, the maximum 0.7~pc distance
between the jet and J1550 corresponds to 7~Mpc for a supermassive
black hole of $10^{8}$~$M_{\odot}$.

In a previous study, \citet{WDL_2003} modeled the evolution and light
curve of J1550's ballistic jets using the same model we employ,
namely, an expanding jet interacting with the interstellar medium
(ISM).  They modeled the data for the eastern jet, attributing the
X-ray emission to a reverse shock, and found that the gas density
around J1550 is unusually low.  Later, their work was extended by
\citet{Hao_Zhang_2009} to include the western jet.  Both groups
focused their attention on the properties of the environment around
J1550; accordingly, they adopted nominal and fixed values for jet
inclination (50$\degr$ and 68$\degr$, respectively), initial Lorentz
factor (3), and jet energy (3.6$\times10^{44}$ erg).  Both groups
found evidence for the existence of a low density cavity around J1550
(modeled in more detail by Hao \& Zhang), and a possible east-west
asymmetry in the ambient gas.
   
While we follow in the footsteps of Wang et al. and Hao \& Zhang, our
aim is different.  We are focused on the question of the alignment of
the inclination angle of the black hole's spin axis and the orbital
inclination angle.  Therefore, in distinction with the earlier work,
we disregard the X-ray light-curve data, which are primarily useful in
constraining the emission mechanisms or the electron density and
magnetic fields in the jet.  Rather, we concentrate on modeling the
kinematics of the ballistic jets and deriving reliable values and
error estimates for the kinematic parameters.  The parameter of chief
interest is the inclination of the black hole's spin axis.

\section{Data}\label{section:data}
 
We use archival \chandra {\em X-ray Observatory} data for eight
observations of J1550 that were obtained using the Advanced CCD
Imaging Spectrometer (ACIS) between 2000 June and 2003 October.  The
exposure times range from $4-50$~ks.  Pipeline processed
level-2 event files\footnote{using CXC DS-7.6.10} were used to produce
images of the field of J1550.  When detected, images of the eastern
(approaching) jet yielded 16--40 counts and the western (receding) jet
100--400 counts; J1550 itself was always detected and yielded 60--3000
counts.

These same \chandra data were used by \citet{Hao_Zhang_2009} in their
analysis of the X-ray jets.  They relied on the absolute astrometric
precision of \chandra in order to derive positions for each jet and
thereby its offset from J1550.  We have reduced the astrometric errors
severalfold by directly measuring in each image the relative
separations between J1550 and the jets.

In measuring the precise jet positions, which are given in
Table~\ref{tab:obs}, we smoothed each image using a 1$\arcsec$
Gaussian kernel and then determined the centroid of each jet using the
DAOphot {\sc find} routine \citep{DAOPHOT}.  This procedure was used
to derive initial estimates for all the jet positions.  Then, 1000
Poisson random realizations of each field were produced, and the
centroid measurements were repeated.  In most cases, the positions for
a given jet were tightly clustered about a single value, and a
separation and error were derived from this distribution.  However,
for three observations of the eastern jet (Obs. X1, X3, and X6 in
Table~\ref{tab:obs}) the images are particularly faint (possibly
because the emission is extended), which resulted in a broad
distribution of positions.  In these cases, a Gaussian-weighted mean
based on the jet position angle $\phi_j$ for each realization $j$ was
used to derive the separation between the jet and J1550.  As a
reference value, we used the average position angle for the jets
$\phi_{\rm PA}$ along with its error $\sigma_{\rm PA}$, $\phi_{\rm PA}
= 94\fdegr25 \pm 0\fdegr3$ (measured east of north).  This value is
consistent with those determined by \citet{Hannikainen_2009} and
\citet{Corbel_2002} and was measured for a single frame generated by
coaligning and coadding all of the X-ray images.  The weights $w_j$
were calculated according to ${\rm log}(w_j) =
-\frac{1}{2}(\phi_j-\phi_{\rm PA})^2/\sigma_{\rm PA}^2$.  Typically,
the position errors for the eastern jet were several tenths of an
arcsec, while for the brighter western jet they were $\lesssim
0.1\arcsec$.

In addition to the positions derived using the \chandra data, we
include in our analysis two radio positions (Obs. R1 and R2 in
Table~\ref{tab:obs}).  These measurements are taken from
\citet{Corbel_2002}, who derived positions from observations obtained
using the Australia Telescope Compact Array (ATCA) on 2000 June 1 and
2002 January 29.  In the first observation, only the eastern jet is
observed, whereas in the second, the eastern jet has faded and the
western jet alone is present.

As a final constraint on our kinematic jet model, we require that the
apparent separation speed of the jets at launch match the value
measured using the LBA, $65.5\pm13.2$ mas/d \citet{Hannikainen_2009}.
This speed and the jet positions are the sole inputs to our principal
model in Section~\ref{section:results}.  However, in
Section~\ref{subsec:radio}, we additionally consider radio intensity
measurements given by \citet{Hannikainen_2009}.  They report 2.29 GHz
flux densities taken four and six days after the X-ray flare with
intensity ratios of $S_{E1}/S_{W1} = 3.55$ and $S_{E2}/S_{W2} = 2.40$,
respectively; we assume that these ratios are uncertain by 25\%.  We
also adopt their measurements of the radio spectral index, $\psi_{1} =
-0.43$ and $\psi_{2} = -0.21$, taken from flux densities measured with
the ACTA at 4.8 and 8.6 GHz.  The spectral index measurements and
corresponding LBA images, while not strictly simultaneous, were
obtained within several hours of one another.

  \begin{deluxetable}{lrcc} 
  \tablecolumns{          4}
  \tablewidth{0pc}  
  \tablecaption{Relative Jet Positions} 
  \tablehead{\colhead{Obs.} & \colhead{$\Delta t^\prime$\tablenotemark{a} (d)} & \colhead{Eastern Offset (arcsec)} & \colhead{Western Offset (arcsec)}}
 \startdata  \label{tab:obs}
R1  &    620.5   &   21.9$\pm$0.3\tablenotemark{b}  &  \nodata  \\
X1  &    628.5   &   21.5$\pm$0.5  &   \nodata  \\
X2  &    701.4   &   22.7$\pm$0.2  &   \nodata \\
X3  &    722.2   &   23.7$\pm$0.5  &   \nodata  \\
R2  &    1227.5 &   \nodata  &             22.6$\pm$0.3\tablenotemark{b}  \\
X4  &    1268.8 &   28.5$\pm$0.2   &  22.78$\pm$0.05 \\
X5  &    1368.5 &   \nodata &                           23.19$\pm$0.07 \\
X6  &    1466.0 &   29.6$\pm$0.6   &  23.44$\pm$0.10 \\
X7  &    1591.3 &   \nodata &                           23.76$\pm$0.10   \\ 
X8  &    1859.6  &   \nodata &                          24.4 $\pm$0.2   \\ 
\enddata
 \tablenotetext{a}{Time since the jets were launched: MJD - 51076.}
 \tablenotetext{b}{ACTA position from \citet{Corbel_2002}. \\ }
\end{deluxetable}

\section{The Jet Model}\label{section:model}

The development of our kinematic jet model follows
\citet{Hao_Zhang_2009} and \citet{WDL_2003}.  The model we use has
been designed to describe gamma-ray bursts, but it is applicable to a
relativistic, adiabatically expanding jet.  To begin, we consider a
pair of symmetric jets, each launched with a kinetic energy $E_0$ and
Lorentz factor $\Gamma_0$.  As the jets expand into their
environments, they entrain material from the surrounding medium,
dissipate their kinetic energy at the shock front and heat the ISM.
We neglect radiative losses and assume that the jets are confined and
evolve adiabatically.  Following \citet{WDL_2003}, we assume that
particles are accelerated uniformly and randomly at the shock front.
Each such jet obeys the relation
\begin{equation}
E_0 = (\Gamma-1) M_0 c^2 + \sigma(\Gamma_{\rm sh}^2-1) m_{\rm sw} c^2,
\label{eq:energy}
\end{equation}
where $\Gamma$ is the instantaneous bulk Lorentz factor of the jet,
$M_0$ is the mass of the jet ejecta, and $\Gamma_{\rm sh}$ is the
Lorentz factor at the shock front.  The mass of the entrained
material, $m_{\rm sw}$, that has been swept up by the shock is
approximately $m_{\rm sw} = \Theta^2 m_{\rm p} n \pi R^3/3 $, where
$\Theta$ and $R$ are respectively the jet half opening angle and the
distance the jet has traveled.  The numerical factor $\sigma$ varies
from $\approx 0.35$ for ultrarelativistic shocks to $\approx 0.73$ in
the nonrelativistic limit \citep{WDL_2003, BM76}.  Following
\citet{Huang_1999}, we adopt a simple numerical scaling to interpolate
between the two regimes: $\sigma = 0.73 - 0.38 \beta$, ($\beta =
\sqrt{1-1/\Gamma^2}$).

At the shock front, the jump condition relates the bulk Lorentz factor
of the jet to that of the shocked gas \citep{BM76}:
\begin{equation}
\Gamma^2_{\rm sh} = \frac{(\Gamma+1)[\hat{\gamma}(\Gamma-1)+1]^2}{\hat{\gamma}(2-\hat{\gamma})(\Gamma-1)+2}.
\end{equation}
The adiabatic index $\hat{\gamma}$ varies between 4/3 and 5/3, which
are respectively its ultrarelativistic and nonrelativistic limits.  We
interpolate between these regimes via $\hat{\gamma} =
(4\Gamma+1)/3\Gamma$ \citep{Huang_1999, WDL_2003, Hao_Zhang_2009}.

On the plane of the sky, the apparent proper motions of the
approaching and receding jets, $\mu_a$ and $\mu_r$, are given by
\begin{equation}
\mu_{a} = \frac{\beta\ c\ {\rm sin } \theta}{D (1-\beta\  {\rm cos} \theta)}, \qquad  \mu_{r} = \frac{\beta\ c\ {\rm sin } \theta}{D (1+\beta\  {\rm cos} \theta)}.
\end{equation}

As we show in Section~\ref{section:results}, the simple model governed
by Eqn.~\ref{eq:energy} fails to fit the observations.  Motivated by
the results of \citeauthor{Hao_Zhang_2009} and \citeauthor{WDL_2003},
we have generalized Eqn.~\ref{eq:energy} to allow for the jets to
first propagate through a low density cavity before encountering and
shocking against the ISM.  In the east-west direction, we allow for
the cavity to differ in size. We additionally consider the
possibility of an intrinsic asymmetry in the jets.
Eqn.~\ref{eq:energy} becomes:
\begin{equation}
\eta E_0 = (\Gamma-1) \eta M_0 c^2 + \sigma(\Gamma_{\rm sh}^2-1) m_{\rm sw} c^2, \\
\label{eq:energyupdate}
\end{equation}
and the entrained mass is now 
\begin{equation}
m_{\rm sw} = \frac{\Theta^2 m_{\rm p} n \pi}{3} \times \left\{ 
	\begin{array}{l l} 
            R^3, & \quad R \leq \zeta R_{\rm cr},\\
		(\zeta R_{\rm cr})^3 + \delta [R^3 - (\zeta R_{\rm cr})^3],  & \quad R > \zeta R_{\rm cr},
	\end{array}  \right.
\label{eq:mupdate}
\end{equation}
where $R_{\rm cr}$ and $\delta$ are respectively the radius of the
cavity centered on J1550 and the density jump at the cavity boundary.
The ratio of the western-to-eastern cavity dimensions is given by
$\zeta$.  Similarly, $\eta \equiv \frac{(E_0/n\Theta^2)_{\rm
    west}}{(E_0/n\Theta^2)_{\rm east}}$ parameterizes the asymmetry of
the jets.  In application, the asymmetry parameters $\zeta$ and $\eta$
are taken to be unity for the eastern jet and can vary for the western
jet.

In order to obtain a model solution for a particular set of
parameters, we evolve the energy equation as the jet expands (either
Eqn.~\ref{eq:energy} or Eqn.~\ref{eq:energyupdate}) in 4-hour
time-steps by sequentially solving for $\Gamma(t)$ in the rest frame
of J1550.  At each time step, we calculate the separation between each
jet and the central source by integrating $\beta(t)$ and by
calculating the projected angles $\alpha$: $\alpha(t^\prime) =
R(t) {\rm sin}~\theta/D$.  Here, $t^\prime = t \mp R(t) {\rm
  cos}~\theta/c$ is the observer's time, which takes into account the
time delay between J1550's rest frame and that of the observer for
whom the light-travel paths of the approaching and receding jets are
respectively shortened and elongated.

Our model requires up to eight physical parameters: $\theta$,
$\Gamma_0$, $D$, $R_{\rm cr}$, $\delta$, $\eta$, $\zeta$, and lastly
the ``effective energy'' $\tilde{E}$ which we now define.  As alluded
to above, a degeneracy exists in our model between jet energy, ambient
gas density, and the jet opening angle.  These three quantities appear
as a single and inseparable term in the kinematic equations, $E_0/ n
\Theta^2$.  To make physical sense of this combined quantity, we
assume that the density of the ISM is a standard $n_{\rm ISM} = 1\
{\rm cm}^{-3}$, so that $n=1/\delta$~cm$^{-3}$, and adopt $\Theta =
1\degr$ (\citealt{Kaaret_2003}).  Predicated upon our assumed values
for $n_{\rm ISM}$ and $\Theta$, the jet energy $E_0$ is then $E_0 =
\tilde{E}$.

Finally, we go beyond our principal, kinematic model to consider the
ratio of the radio intensities of the two jets.  We consider the
simplest case of the ejection of a pair of identical and unimpeded
condensations.  When measured at equal separation from the black hole,
one has 
\begin{equation}
\frac{S_a}{S_r} = \left(\frac{1+\beta\ {\rm cos}\ \theta}{1-\beta\ {\rm cos}\ \theta}\right)^{3-\psi},
\label{eq:radio}
\end{equation}
where $\psi$ is the spectral index and the subscripts $a$ and $r$
refer to the approaching and receding jets, which are taken to be
discrete ejecta \citep{Mirabel_Rodriguez}.  Because the jets are
observed at unequal distances from the black hole, we must adopt a
model of how jet intensity varies with time; we assume a simple
power-law dependence.  Then, allowing for our case of intrinsically
asymmetric jets, Eqn.~\ref{eq:radio} becomes
\begin{equation}
\frac{S_a}{S_r} = \left[\frac{\Gamma_r(1+\beta_r\ {\rm cos}~\theta)}{\Gamma_a(1-\beta_a\ {\rm cos}~\theta)}\right]^{3-\psi-\Delta} \eta^q,
\label{eq:radioupdate}
\end{equation}
where $\Delta$ is a fit parameter, which for positive values describes
a decay in brightness with time.  The effect of jet asymmetry on the
radio emission is captured by $q$, which can range from -1 to 1.5
depending on the source of asymmetry: $E_0$ ($q=-1$), $n$ ($q=0$), or
$\Theta$ ($q\in[1,1.5]$).

\section{Markov Chain Monte Carlo}\label{section:mcmc}

Markov Chain Monte Carlo (MCMC) is a powerful statistical technique by
which random samples are drawn from a posterior distribution of
arbitrary form.  In our case, the posterior distribution is the
probability of our model parameters, given the data.  MCMC algorithms
perform a ``guided walk'' of transitions through parameter space such
that, after an initial burn-in phase, the chain directly reproduces
the likelihood surface for the model.  MCMC has several advantages
over traditional gridded-search algorithms when the number of
parameters is large.  For example, the search time with MCMC scales
approximately linearly with the number of parameters rather than
exponentially \citep{Martinez_2009}.  Furthermore, the ergodic
property of the Markov Chain guarantees (asymptotically) that the
chain will fully explore parameter space and reach the optimum global
solution.

Transitions in the chain are effected via a ``jump''
distribution\footnote{We implement a particular class of the algorithm
  known as random-walk MCMC.  In this approach, a sequence of
  transitions from the current parameter values are proposed and are
  then incrementally accepted or rejected.}  $J(x^* | x_n)$ (e.g., a
multivariate Gaussian) that defines a probability of selecting a
candidate transition to a new state $x^*$ given the current state
$x_n$.  The transition probability from $x_n$ to $x^*$ is governed by
the Metropolis-Hastings algorithm \citep{MH} and is determined by the
ratio $r$ of probability densities 
\begin{equation}
r   =  \frac{ p(x^* | y ) J(x_n | x^*) }{ p(x_n | y) J(x^* | x_n) }, 
\label{eqn:MH}
\end{equation}
where $y$ refers to the data, and $p(a | b)$ should be read in the
usual way as the probability of $a$ given $b$.  The term $p(x^* |
y)/p(x_n | y)$ in the equation above gives the likelihood ratio of the
two states, while the remaining term corrects for bias introduced by
the jump-distribution density at each state.  The state of the next
link in the chain, $x_{n+1}$, is then chosen according to
\begin{equation}
x_{n+1} = \left\{ 
	\begin{array}{l l}	
		x^*, & \quad {\rm with~probability~min}[r,1],\\ 
		x_n, & \quad {\rm otherwise }.
	\end{array}  \right.
\label{eqn:select}
\end{equation}
The likelihood ratio appearing in Eqn.~\ref{eqn:MH} is calculated by
evaluating the $\chi^2$ for each state while taking into account the
prior $\wp$ on all of the model parameters.  In this case, the priors
are introduced independently so that $\wp \equiv
\displaystyle\prod_{k=1}^{\rm N} \wp_k$, where N is the number of
parameters and $\wp_k$ gives the prior for parameter $k$.  Omitting
additive constants, the log-likelihood for state $x$ is
\begin{equation}
{\rm log}\left(p(x | y)\right)  =  -\frac{1}{2}\left[\chi^2(x) - 2~ {\rm log}\left(\wp(x)\right)\right].
\label{eqn:lik}
\end{equation}

\subsection{MCMC in Practice}

As Eqn.~\ref{eqn:lik} makes apparent, the prior acts as a penalty to
$\chi^2$, and for the special case that a prior is ``flat'' (i.e.,
independent of $x$), one recovers the usual least-squares formula.  It
is also worth noting that because the prior only enters into the MCMC
chain generation as a ratio (Eqn.~\ref{eqn:MH}) the scaling of the
prior is arbitrary.  We introduce a new term for this penalized
$\chi^2$, namely $\chi^2_\wp$, such that 
\begin{equation}
\chi^2_\wp(x)\equiv\chi^2(x) - 2~ {\rm log}\left(\wp(x)\right).  
\end{equation}
Unless stated otherwise, we choose to normalize the prior so that the
penalty term (2~log$\left[\wp(x)\right]$) is zero at the best fit,
i.e., at the minimum value of $\chi^2_\wp$.

We adopt an asymmetric Gaussian prior on the distance because it has
been previously measured using optical and near-infrared photometry
(see Section~\ref{section:Intro}; \citealt{Orosz_Steiner_2011}).  For
the asymmetric terms, we adopt a log-flat prior on the difference from
unity, i.e., $\wp_\eta \propto {\rm min}[1/\eta,\eta]$ (and likewise
for $\wp_\zeta$).  As an example and stated differently, we consider a
term implying a 10-fold asymmetry to be {\em a priori} one tenth as
likely as one that is symmetric.  We adopt flat priors on $\theta$ and
$R_{\rm cr}$ and flat priors on the log-values of scale parameters
(i.e., the jet energy, $\Gamma_0$ and $\delta$).  The priors and
parameter ranges\footnote{While it is optimal to use an unbounded
parameter space in performing MCMC sampling, it is also sensible to
set physically meaningful constraints on the parameters (e.g.,
$\Gamma_0 > 1$).  To achieve both objectives, we have transformed each
parameter using a logit function to map a parameter $z$ from its range
[$z_{\rm min},z_{\rm max}$] onto an infinite scale: ${\rm logit}(t)
\equiv z_{\rm min} + ({z_{\rm max}-z_{\rm min}})/({1+e^{-t}})$ for
$-\infty<t<\infty$.} are discussed further and illustrated in
Section~\ref{section:results}.

In order to initialize the chain and the jump distribution, we make
starting guesses for the model parameters and step sizes.  These
initial values are improved upon by running a sequence of ``training''
iterations.  The training phase incrementally improves the jump
function until its shape is a close approximation to the posterior
covariance matrix, thereby greatly increasing the MCMC efficiency.
The sequence becomes increasingly tuned to the likelihood surface,
simultaneously refining $\Sigma$ (the covariance
estimate)\footnote{$\Sigma$ is calculated from the chain positions and
  is used to define the jump function for each sequence.  The jump
  function is taken to be a $t-$distribution with 4 degrees of freedom
  that is symmetric about the present position.} and optimizing the
solution.

The training phase continued for a minimum of 15 iterations, each of
which generated a trial chain with 2000 elements. Training terminated
either after 25 cycles were completed or when the chain attained an
acceptance fraction between 24\% and 37\%\footnote{The target
acceptance fraction was set at $\approx 32\%$.  The optimal value
ranges from $\approx 23\%$ for an infinite-dimensional problem to
$\approx 45\%$ for a univariate problem \citep{Gelman95}. Each run
produced an acceptance fraction of at least 20\%}.

Upon completing the training cycle, 8 chains were generated and run in
parallel using the trained jump function, each to a length of 110
thousand elements.  Seven of the starting positions were chosen by
sampling using a dispersed covariance $\Sigma^\prime=10~\Sigma$ about
the final training position, and the eighth was started directly from
the end location reached by the training sequence.  The initial 10,000
elements of each chain were rejected as the ``burn-in'' phase during
which the chains relax toward a stationary distribution.  Our final
results are based on a total of 8$\times10^5$ MCMC samples.
Convergence of the MCMC run is determined using the criterion of
\citet{Gelman_Rubin}, $\hat{R}$.  The closeness of this criterion to
unity is the measure of convergence.

\begin{figure}
{\includegraphics[clip=true, angle=90,width=8.85cm]{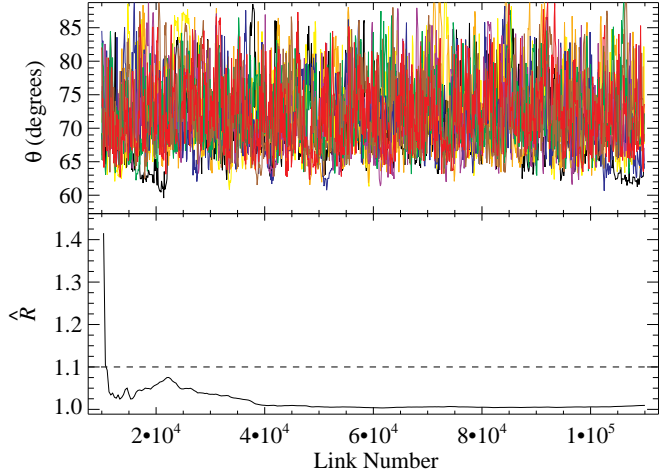}}
\caption{{\it top}: The trace of $\theta$ for Model AC of
  Section~\ref{section:results}.  Eight parallel chains are used; for
  each, the initial $10^4$ elements are generated during the burn-in
  phase and discarded from the analysis.  {\it bottom}: The
  convergence of the chain over time.  The chains reach convergence
  quickly, which is indicative of efficient sampling.}\label{fig:convergence}
\end{figure}

In Figure~\ref{fig:convergence}, we plot a trace of our parallel runs
over time for inclination in our adopted model (see
Section~\ref{section:results}).  In the bottom panel, we show the
Gelman \& Rubin convergence diagnostic of the chain over time.
Typically, a chain is considered converged if $\hat{R}\leq1.1$, or 1.2
(see, e.g., \citealt{Verde_2003})\footnote{Larger values of $\hat{R}$
suggest that either the parameter space is insufficiently sampled or
that the chains are not fully evolved.}.  For $\theta$, our parameter
of interest, we obtain $\hat{R} < 1.01$.

\begin{figure}
{\includegraphics[clip=true, angle=90,width=8.85cm]{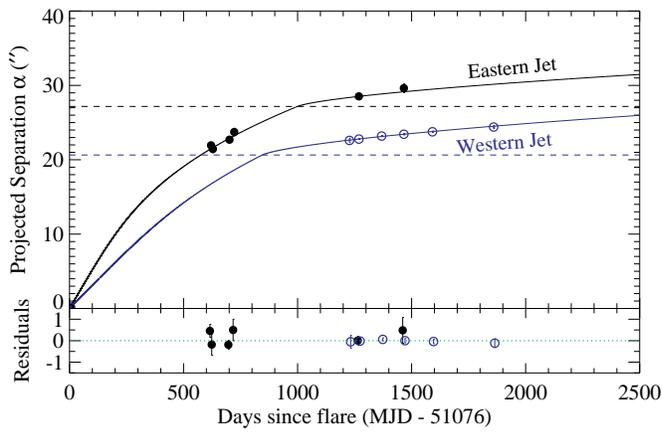}}
\caption{The best-fitting model and fit residuals for the eastern jet
  (filled circles) and western jet (open circles).  The cavity
  locations are marked by dashed horizontal lines, which indicate that
  the western wall (for the receding jet) is closer to the black hole
  than the eastern wall.  For clarity, residuals for the
  coincidentally detected eastern and western jets are shown slightly
  offset in time.  In the top panel, the error bars are smaller than
  the symbols.}\label{fig:fit}
\end{figure}

\begin{figure}
{\includegraphics[clip=true, angle=0,width=8.85cm]{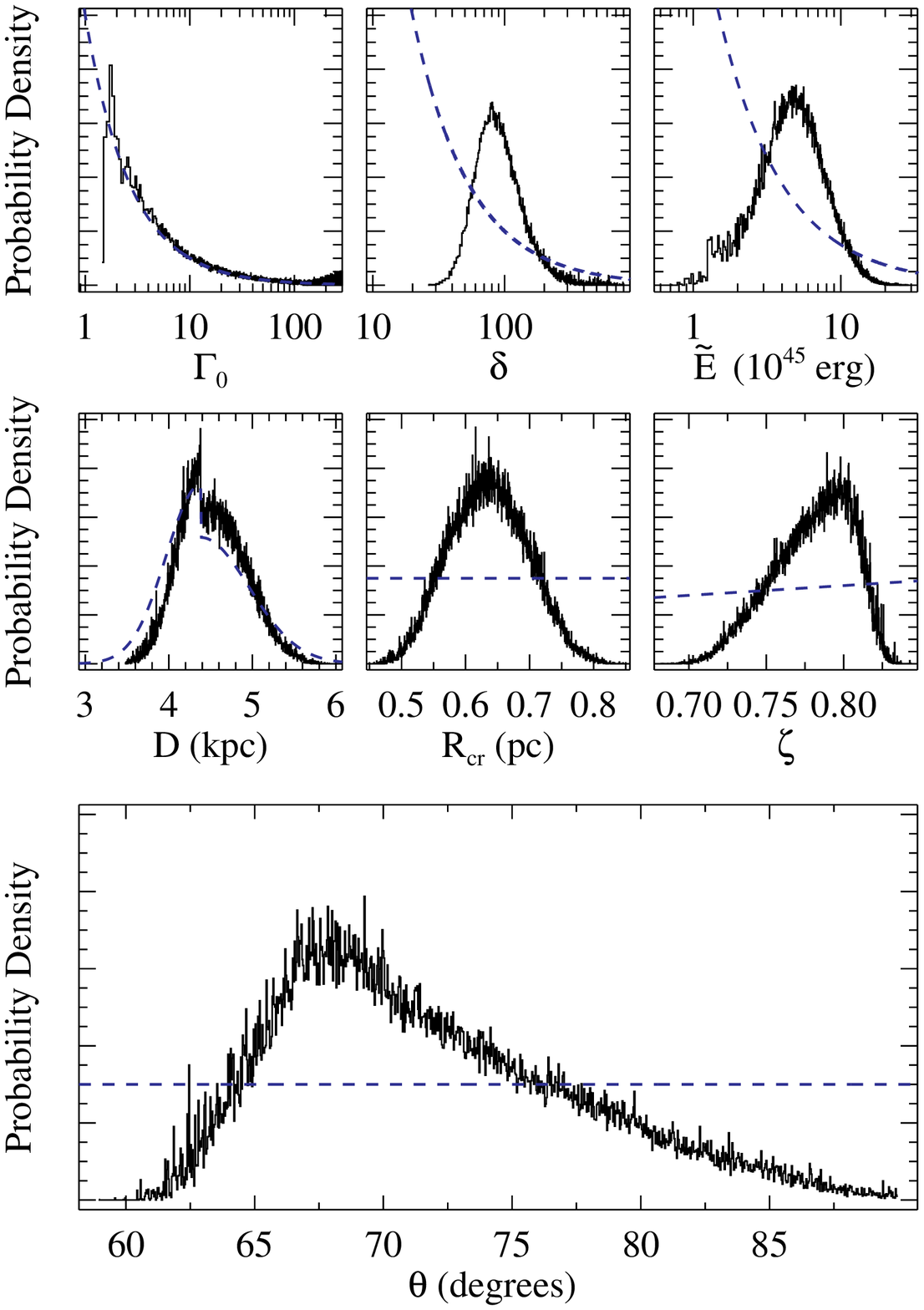}}
\caption{MCMC results for Model AC.  Probability densities are shown
  for each parameter on an arbitrary scale and have been obtained by
  marginalizing over all other parameters.  An overlay for each prior
  shape is drawn as a dashed line.  Note that the only two
  parameters which closely track the prior function are the system
  distance and $\Gamma_0$ (at high values only).  Otherwise, the prior
  contributes minimally to the parameter distribution.
}\label{fig:mcmcresults}
\end{figure}

\section{Results}\label{section:results}

In this section we consider three symmetric-jet models, including our
adopted model.  For these models, and for the additional models
discussed in the following section, we assume that the jets were
launched at the time of J1550's giant X-ray flare
(Section~\ref{section:Intro}).

\subsection{Two Preliminary Models}

We first consider and rule out two simple models.  For the simpler of
these, which we refer to as Model S1, the jets are symmetric and
propagate through a uniform medium (Eqn.~\ref{eq:energy}; i.e., $\eta =
\zeta = \delta =1$ and $R_{\rm cr}=0$).  The strong deceleration of
the jets at late times is not accommodated by this model, and the best
fit achieved is unacceptable, $\pchinu = 68$.  For Model S2, we
introduce a symmetric cavity centered on J1550 with $\delta$
and $R_{\rm cr}$ as free fit parameters.  The fit is significantly
improved, $\pchinu = 42$, but it is still far from acceptable.  The
results for both models are given in Table~\ref{tab:results}.

\subsection{Our Adopted Model} 

We now consider our primary model -- an extension of Model~S2 that
allows the source to be positioned off-center in the cavity.  This
asymmetric cavity model (Model AC) is obtained by freeing the fit
parameter $\zeta$ (while leaving $\eta$ fixed at unity; see
Eqn.~\ref{eq:energyupdate}).  As illustrated in Figure~\ref{fig:fit},
for a modest (22\%) degree of asymmetry, this model produces a
successful fit to the data with $\pchinu = 1.44$.  Results are given
in Table~\ref{tab:results} and marginal distributions from the MCMC
run are shown for each parameter in Figure~\ref{fig:mcmcresults}.

The eastern and western cavity walls are located respectively at
$0.6$\ pc and $0.5$\ pc from the black hole and the density contrast
at the boundary of the cavity is $\sim100$.  The gas density within
the cavity is much lower than that of the ISM.  This must be the case
in order for the jets to have passed through without sweeping up
enough mass to halt their expansion.  Motion within the cavity lasted
for $\approx$1.5 years (in the frame of J1550), until the receding
western jet impacted the dense ISM at its cavity wall and abruptly
began decelerating in advance of its eastern counterpart (see
Fig.~\ref{fig:fit}).

\begin{figure}
{\includegraphics[clip=true, angle=90,width=8.85cm]{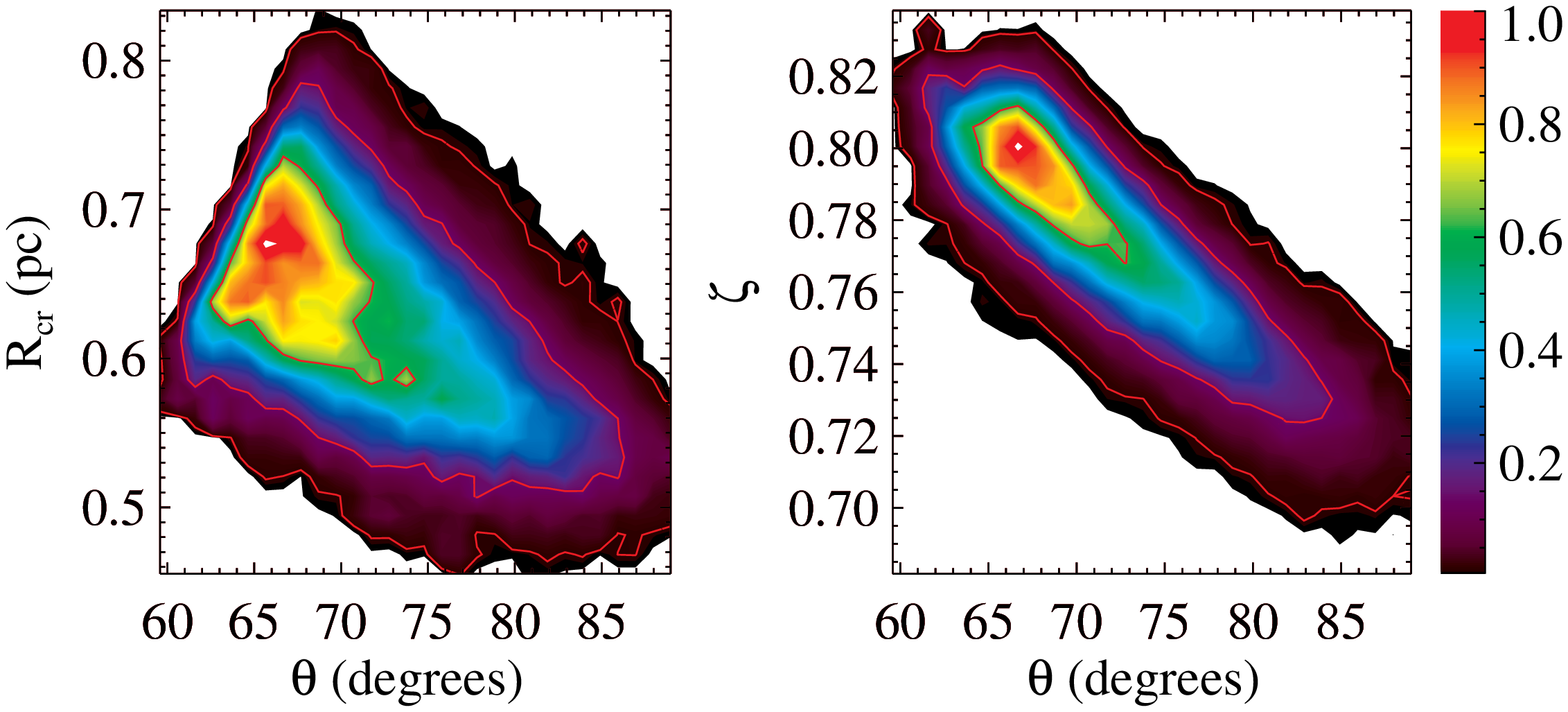}}
\caption{Shown for Model AC are the MCMC density contours for the two
  parameters that correlate most strongly with inclination: cavity
  size $R_{\rm cr}$ and the cavity asymmetry parameter $\zeta$.  The
  densities are calculated by marginalizing over all unshown
  parameters.  Red contours mark the 68\%, 95\%, and 99.7\% confidence
  regions about the most likely value, which is normalized to a
  density of unity.  The central value of $\theta$ changes from
  70$\degr$ to 80$\degr$ as $R_{\rm cr}$ varies from 0.65 pc to 0.58
  pc, and as $\zeta$ decreases from 0.78 to 0.74.
}\label{fig:contours}
\end{figure}

The total energy for both jets is an impressive $E_{\rm tot} \approx
10^{46}~{\rm erg}~ \frac{n_{\rm ISM}}{1 \cm^{-3}}
\left(\frac{\Theta}{1~\rm{deg}}\right)^2$.  At launch, the Lorentz
factor of the jets is constrained to be $\Gamma_0 > 1.6$ (99.7\%
confidence).  However, the data provide no upper limit on $\Gamma$, as
implied by Figure~\ref{fig:mcmcresults}, which shows that for large
values of $\Gamma$ the distribution closely tracks the prior.
Likewise, the data only weakly constrain J1550's distance.  However,
the remaining five parameters are well determined by the data and are
quite independent of their priors (Fig.~\ref{fig:mcmcresults}).  For
the key parameter, the jet inclination angle, we obtain $\theta
\approx 71\degr$ ($64\degr < \theta < 83\degr$ at 90\% confidence) and
find only moderate correlations with the other fit parameters.  The
strongest of these correlations are with $\zeta$ and with $R_{\rm
  cr}$, which are illustrated in Figure~\ref{fig:contours}.

\subsection{Constraining Spin-Orbit Alignment}

We now use Model AC and the results of our MCMC analysis to examine
the relationship between the spin axis of the black hole (the same as
that of the jet; see Section~\ref{section:Intro}), and the orbital
angular momentum vector.  We assume that the inclination of the
orbital plane $i$ is Gaussian distributed: $i=74\fdegr7\pm3\fdegr8$
\citep{Orosz_Steiner_2011}.

\begin{figure}
{\includegraphics[clip=true, angle=90,width=8.85cm]{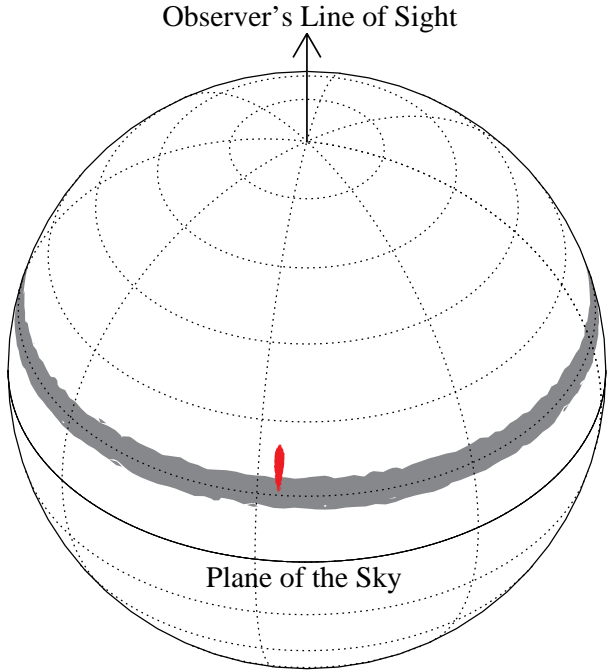}}
\caption{ The celestial sphere centered about J1550 with the observer
  situated along the pole.  The angular momentum of the orbital plane
  is constrained to lie along the grey band (drawn with 1$\sigma$
  width), and the spin angular momentum axis derived from the jets is
  overlaid in red (1$\sigma$ about the most likely value).  The
  position angle is completely unbounded for the orbital angular
  momentum, whereas the jets provide a tight constraint on the
  position angle of the black hole's spin axis
  (Section~\ref{section:data}).  In fact, the uncertainty in the
  position angle of the jets is so small ($\pm0.3$~deg) that for
  purposes of illustration it has been tripled to make it visible in
  this figure.}\label{fig:globe}
\end{figure}

\begin{figure}
{\includegraphics[clip=true, angle=90,width=8.85cm]{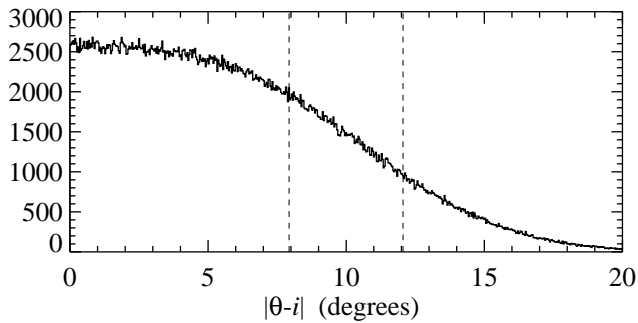}}
\caption{The difference in orbital and jet inclination angles derived
  from the MCMC run of Model AC.  The results show no sign of a
  misalignment along the line of sight; 68\% and 90\% upper limits on the difference between
  inclinations are $8\degr$ and 12$\degr$, respectively.
}\label{fig:tilt}
\end{figure}

Our constraints on the locations of both axes are illustrated in
Figure~\ref{fig:globe}, which was derived using one million
Monte-Carlo draws to represent each axis.  This figure shows how
readily our results are able to falsify the alignment hypothesis, even
though we lack a measurement of the position angle of the orbital
plane.  Specifically, (1) over 80\% of the sky, we are able to rule
out the possibility that the spin and orbital axes are
aligned\footnote{at 90\% confidence}; and (2) the probability by
random chance that the inclination angles agree so closely (see
Fig.~\ref{fig:tilt}) is less than 10\%.

Because the continuum-fitting method depends only on inclination angle
(and not position angle), and because the difference between the
inclination $\theta$ of the spin/jet axis and the inclination $i$ of
the orbital plane is of critical importance in measuring the spin of
J1550, we now use Model AC to determine $\theta - i$.  Our results are
shown in Figure~\ref{fig:tilt} where it is obvious that there is no
evidence for any misalignment along the line of sight.  That is, our results are consistent
with $\theta=i$.  We place upper limits on the absolute difference
between orbital and spin inclinations of $8\degr$ and $12\degr$ at the
68\% and 90\% levels of confidence, respectively.

Given the $<10$\% a priori chance that the inclination angles agree as
closely as measured, our results provide support for the hypothesis
that the two axes are aligned.  However, without knowledge of the
position angle of the binary axis (which can lie anywhere along the
grey band in Fig.~\ref{fig:globe}), we cannot conclude whether they
are, in fact, aligned.


  \begin{deluxetable}{lccccccccccc} 
  \tablecolumns{          11}
  \tablewidth{0pc}  
  \tablecaption{Kinematic Model Settings and Fit Results} 
  \tablehead{\colhead{Parameter} & \colhead{Range} & \colhead{Prior Shape\tablenotemark{a}}  & \colhead{Model S1} & \colhead{Model S2} &  \colhead{Model AC}  }
 \startdata  \label{tab:results}
 $\theta$ (degrees)     &    $0 - 89.99$  &  F                                                    &  $53.9\pm0.7$                   & $58.16\pm1.7$                        &   $70.8_{-4.5}^{+7.3}$ \\    
 \hline
 $\Gamma_0$              &    $1 - 1000$  & LF                                                   &   $210^{+390}_{-160}$      & $50_{-43}^{+320}$                &     $36_{-32}^{+300}$ \\  
 $\tilde{E}\tablenotemark{b} $ ($10^{45}$ erg)     & $10^{-10} - 10^{10}$ & LF                           &   $91.8^{+9.6}_{-6.7}  $     & $ 74^{+18}_{-14} $               &     $  5.9_{-2.3}^{+3.6}$  \\ 
 $D $ (kpc)                   &    $3 - 7$\tablenotemark{c}&  N(4.38$^{+0.58}_{-0.41}$)    &   $3.07\pm0.06$     & $4.30_{-0.23}^{+0.29}$         &     $ 4.48_{-0.34}^{+0.43}$ \\   
 $R_{\rm cr}$ (pc)         &  $0 - 5$   & LF                                                          &  \nodata                             & $0.46\pm0.03$                      &    $  0.63\pm0.06$\\        
 $\delta $                       &  $0.1 - 10^4$  & LF                                                  &  \nodata                             & $940_{-790}^{+4900}$          &      $104_{-34}^{+70} $ \\ 
 $\zeta $                        &  $10^{-2} - 10^2$ & LF (max[$\zeta$,$\zeta^{-1}$]) &  \nodata  &                       \nodata                                       &     $ 0.78\pm0.03$  \\
 \hline
min($\pchinu$)                       &   \nodata  & \nodata                                           & 67.93 (543.4/8)               &    41.59 (249.6/6)                   &   1.44 (7.21/5)            \\
min($\rchinu$)                       &   \nodata  & \nodata                                            & 67.61                               &    40.94                                  &   1.22                         
\enddata
\tablecomments{The values quoted are the median parameter and symmetric 68\% confidence interval (1$\sigma$ equivalent) derived from the MCMC run (as opposed to the single best-fit values).  }
 \tablenotetext{a}{F is flat, LF is log-flat, and N is a normal distribution.}
 \tablenotetext{b}{Assumes $n_{\rm ISM}=1$~cm$^{-3}$ and $\Theta=1\degr$.}
 \tablenotetext{c}{The lower bound on distance is taken from
 \citet{Hannikainen_2009} and the upper bound is derived using $D \leq \frac{c}{\sqrt{\mu_a\mu_r}}$ \citep{Mirabel_Rodriguez}.}
 \end{deluxetable}

\section{Radio Intensities and Asymmetric-Jet Models}\label{subsec:radio}

We now consider the radio intensity measurements discussed in
Section~\ref{section:data} in order (1) to identify any intrinsic
asymmetry in the jets and (2) to check the consistency of our
kinematic model.  In doing this, we are motivated by observations of
the microquasar GRO J1655--40, which in 1994 displayed multiple
ejection events, each of which expanded and decayed on a time scale of
a few days.  The approaching and receding jets were found to be
intrinsically asymmetric; additionally, the sense of the asymmetry
changed from event to event \citep{Hjellming_Rupen, Mirabel_Rodriguez}.

Before introducing intrinsic jet asymmetry into the model, we first
proceed by extending Model AC to create Model RAC.  This latter model
retains the traits of Model AC but now incorporates
Eqn.~\ref{eq:radioupdate} and uses the additional free parameter
$\Delta$ to model the radio data.  The data set for Model RAC is
likewise extended and includes its two radio intensity measurements
(Section~\ref{section:data}; \citealt{Hannikainen_2009}).  Re-fitting
the data using Model RAC and comparing with the results obtained for
Model AC, we find a slight ($2\degr$) increase in the jet angle and
similar small changes in the other parameters
(Table~\ref{tab:moreresults}).  The fit is good, $\pchinu = 1.44$, and
$\Delta$, the decay rate of the jet emission, is positive and in the
range $\approx1-5$.

We now examine intrinsic jet asymmetry, and introduce Model RAJ, a
model that considers both kinematics and radio emission.  In this
case, the cavity is presumed to be symmetric, while the energy term
($E_0/n\Theta^2$) is allowed to vary between the eastern and western
jets.  Specifically, we set $\zeta = 1$, free $\eta$, and introduce
the parameter $q$, which characterizes the type of asymmetry in the
jets (Eqn.~\ref{eq:radioupdate}).  The fit results for Models RAC and
RAJ are shown in Table~\ref{tab:moreresults}.  Model RAJ returns a
significantly larger jet inclination angle than Model RAC, $\theta
\approx 82\degr$, and it implies a large difference between the
eastern and western jets, $\eta^{-1} \sim15$.  Because $q\approx0$,
for this model the gross asymmetry can be attributed to an east-west
difference in the gas density (rather than an asymmetry in the
energies or opening angles of the jets; see
Section~\ref{section:model}).

To assess the performance of Model RAC relative to Model RAJ, we
exploit the similarities in the way these models are structured.  In
particular, their respective priors have identical form.  Therefore,
because we attribute equal likelihood to either type of asymmetry, we
can apply the penalty normalization from Model RAC to Model RAJ.  This
yields the goodness-of-fit results shown in
Table~\ref{tab:moreresults}.  Model RAJ is effectively ruled out:
min($\chi2_{\wp, {\rm RAJ}}) - $min($\chi2_{\wp, {\rm RAC}}) = 8.2$.

We now test the strength of this result by considering a
kinematic-only variant of this asymmetric-jet model, Model AJ, which
ignores the radio intensity data.  Model AJ has the virtue that it can
be directly compared with our primary model, Model AC, because both
models have the same number of parameters (seven) and their priors are
identically structured.  As in the comparison above, we apply the
penalty normalization of Model AC (Section~\ref{section:results}) to
Model AJ.  The fit results for the two models are given respectively
in Tables~\ref{tab:results} and \ref{tab:moreresults}.  Based on the
substantial difference in $\chi2_\wp$, min($\chi2_{\wp, {\rm AJ}}) -
$min($\chi2_{\wp, {\rm AC}}) = 7.5$, and the even larger difference
obtained when the radio-intensity data are included, we conclude that
the asymmetric cavity model is favored over the asymmetric jet model
at the $99\%$ level of confidence.

Unlike the manifestly asymmetric jets of GRO J1655--40, the available
evidence indicates that the jets of J1550 are likely intrinsically
symmetric: Model AC is favored over Model AJ, and Model RAJ implies an
implausibly large (factor of 15) difference in the density of the ISM
from west to east.

In comparison with Model AJ or RAJ, our adopted Model AC gives a
reasonable and satisfying description of J1550 as a system comprised
of intrinsically symmetric jets propagating through an evacuated
cavity with eastern and western walls located out at 0.6 pc and 0.5
pc, respectively.  


   \begin{deluxetable}{lccccccccccc} 
  \tablecolumns{          11}
  \tablewidth{0pc}  
  \tablecaption{Additional Model Results}
  \tablehead{\colhead{Parameter} &    \colhead{Model  AJ}  & \colhead{Model  RAC}  & \colhead{Model RAJ}  }
 \startdata  \label{tab:moreresults}
 $\theta$ (degrees)                 &   $86.2_{-3.1}^{+2.4}$      & $72.8_{-5.4}^{+7.4}$          &      $81.9_{-6.8}^{+5.1} $  \\
\hline
 $\Gamma_0$                         &   $22_{-19}^{+270}$         & $37_{-33}^{+390}$          &      $1.41_{-0.14}^{+0.33} $  \\
 $\tilde{E}\tablenotemark{a} $ ($10^{45}$ erg)                     & $213_{-65}^{+83}$          & $6.1_{-2.3}^{+3.8}$          &      $80_{-34}^{+30} $  \\
 $D $ (kpc)                               & $4.83\pm0.36$                & $4.49_{-0.35}^{+0.43}$    &      $3.57_{-0.44}^{+0.50} $\\
 $R_{\rm cr}$ (pc)                     & $0.46\pm0.05$               & $0.63\pm 0.06$                 &      $0.35_{-0.05}^{+0.04} $ \\
 $\delta $                                  & $510_{-410}^{+1700}$    & $98_{-30}^{+57}$            &      $740_{-590}^{+3300} $  & \\
 $\zeta $                        	       & \nodata                            & $0.78\pm0.03$               &      \nodata               \\
 $\eta  $\tablenotemark{b}        & $0.065\pm 0.014$           &  \nodata                            &      $0.068_{-0.013}^{+0.016}$     \\
 $\Delta$\tablenotemark{c}     &  \nodata                             &  $1.9_{-1.1}^{+3.2}$          &      $1.8_{-6.5}^{+5.3} $  \\
 $q$\tablenotemark{d}            &  \nodata                             &  \nodata                            &      $-0.28_{-0.35}^{+0.52} $ \\
\hline
min($\pchinu$)                       &   2.95 (14.74/5)\tablenotemark{c}                  & 1.44 (8.63/6)        &    3.36 (16.81/5)\tablenotemark{c} \\ 
min($\rchinu$)                       &    1.11                                 & 1.11                            &    1.31            
\enddata
\tablecomments{The values quoted are the median parameter and symmetric 68\% confidence interval (1$\sigma$ equivalent) derived from the MCMC run.}
 \tablenotetext{a}{Assumes $n_{\rm ISM}=1$~cm$^{-3}$ and $\Theta=1\degr$.}
 \tablenotetext{b}{The form of the prior for $\zeta$ and $\eta$ are identical (see Table~\ref{tab:results}). }
 \tablenotetext{c}{A flat prior is used for both $\Delta$ and $q$.  The former is allowed to take values between [-10,10] and the latter is constrained to the range [-1,1.5].}
 \tablenotetext{d}{The penalty normalization for Model AJ is taken from Model AC; likewise that for Model RAJ is from Model RAC.} 
 \end{deluxetable}

\section{Discussion}\label{section:discussion}

If we assume that the jets were produced continuously over the
day-long Eddington-limited X-ray flare \citep{Steiner_j1550spin_2011},
then the nominal total jet energy of $\approx 10^{46}$ erg implies that a
significant fraction of the mass accreted onto J1550 during the flare
was directly used to fuel the jets.  Roughly, the initial mass in the
jets was then $\sim 10^{24}$~g and the matter was accelerated to
$\Gamma_0\sim10$.  

We note that the moderate asymmetry we find (with the western cavity
$\approx20$\% smaller in radius than the eastern) is opposite in sense
from the asymmetry determined by \citet{Hao_Zhang_2009}.  We attribute
this difference to several factors: Hao \& Zhang simply adopted
reasonable, ad-hoc values for several key parameters ($\theta$,
$\tilde{E}$, and $\Gamma_0$), and they found a high degree of
asymmetry with $\eta^{-1} \approx 30$ and $\zeta = 1.4$.  (We note
that this particular pair of values of $\eta$ and $\zeta$ allowed a
reasonable fit to be achieved to their data set.)  By improving the
quality and quantity of the astrometric data, we were able to
determine that just one asymmetry parameter is required to explain the
data, and that the resultant asymmetry is less extreme.

Based on results obtained for the sub-pc scale ($\lesssim 0.1$ pc)
jets of GRS 1915+105 and GRO J1655--40, \citet{Heinz_2002} has
proposed that black hole microquasars preferentially inhabit
environments that are under-dense compared to their AGN counterparts.
Heinz offers several explanations, notably that microquasars may
produce self-encasing low density bubbles either as a remnant of the
birthing supernova explosion, or via persistent kinetic outflows from
the compact source.

The enthalpy of the low density cavity in J1550, $\sim10^{40}-10^{42}$
erg, is likely maintained by the steady (or quasi-steady) AU-scale
jets known to be present in the hard or quiescent state of black hole
binaries \citep{RM06, Gallo_2006}.  The $\sim20\%$ measured asymmetry
in the east-west extent of the cavity is unlikely to be a result of a
high proper motion of the binary because this would require an extreme
velocity $\sim 0.1c$.  Rather, this asymmetry is easily explained as
arising from a moderate 20\% variation in the density of the ISM
across the pc-scale region spanned by the jets.  This supposition is
quite plausible, given that J1550 is located only $\sim140$ pc from
the Galactic plane.

One interesting feature of our best-fitting model is shown in
Figure~\ref{fig:fit}: The onset of X-ray emission for the western jet
is first observed after the jet has reached the outer wall of the
cavity, whereas for the eastern jet it occurs well before reaching the
outer wall.  Although there is not enough data to draw a firm
conclusion, this difference in behavior suggests that our model
oversimplifies by describing a succession of low-grade density jumps
(from previous episodes of jet activity) as one single jump at $R_{\rm
  cr}$.  Alternatively, perhaps one or several dense filaments of gas
breached the eastern cavity walls, causing X-ray brightening at the
shock front, but without contributing appreciable mass.

We close our discussion by noting again that our lack of knowledge of
the position angle of the binary restricts us to testing for
spin-orbit alignment along the line of sight.  The test we have
performed nevertheless provides important support for the
continuum-fitting measurement of J1550's spin, which used the orbital
inclination angle as a proxy for the inclination of the black-hole
spin axis \citep{Steiner_j1550spin_2011}.  For the case of J1550, we
have shown that these two inclination angles are consistent within
several degrees.

\section{Conclusions}

Building on earlier work by \citet{Hao_Zhang_2009} and
\citet{WDL_2003}, we have used \chandra and radio imaging data to
model the ballistic motion of the jets of J1550.  We take the time of
J1550's giant X-ray flare, which was promptly accompanied by the
ejection of small-scale ($\sim1000$~AU) relativistic radio jets, as
the launch date of the large-scale ballistic jets.  Using our MCMC
code and a kinematic model of the jets, we find that J1550 is enclosed
in a pc-scale cavity that is moderately asymmetric, and that the jets
are inclined by between 64$\degr$ and 83$\degr$ to our line of sight
(90\% confidence).  These impulsive jets are extremely energetic,
having been launched with a total energy of $\sim10^{46}~{\rm erg}~
\frac{n_{\rm ISM}}{1 \cm^{-3}}
\left(\frac{\Theta}{1~\rm{deg}}\right)^2$.

By comparing our derived inclination angle for the spin axis of the
black hole (taken to be the jet inclination angle) to the orbital
inclination angle, we arrived at our primary result: We find no
evidence for misalignment in our comparison of orbital and jet
inclinations, and we conclude that the spin and orbital inclinations
differ by $<12$~degrees (90\% confidence).  This result has a
likelihood of less than 10\% of occurring by chance.

Theory predicts that accretion torques acting over time will have
brought most black holes into alignment with the orbital plane of
their binary hosts.  This prediction underpins the continuum-fitting
method.  In the case of J1550, our results provide support for such
alignment and for the measured spin of its black hole primary.

\acknowledgements

JFS was supported by the Smithsonian Institution Endowment Funds and
JEM acknowledges support from NASA grant NNX11AD08G.  We thank Bob
Penna, Sasha Tchekovskoy, and Ramesh Narayan for constructive ideas
which helped shaped the direction of this work, and both Joey Neilsen
and Lijun Gou for their comments on the manuscript.  We thank an
anonymous referee for helpful feedback which has improved this paper.
The MCMC analyses were run using the Odyssey cluster supported by the
FAS Science Division Research Computing Group at Harvard University.

\newcounter{BIBcounter}        
\refstepcounter{BIBcounter}

\end{document}